\documentclass[preprint,showpacs,amsfonts,aps,prc,superscriptaddress,nofootinbib,floatfix]{revtex4-1}
\usepackage{amsfonts}
\usepackage{amsthm}
\usepackage{bm}
\usepackage{graphicx}
\usepackage{rotating}
\usepackage{amsmath}
\usepackage[usenames,dvipsnames]{color}
\usepackage{soul}
\usepackage{hyperref}
\usepackage{color}
\usepackage{subfigure}

\def\eq{{\,=\,}}

\begin{document}
\setstcolor{red}
\bibliographystyle{unsrt}


\title{Hydrodynamic flow amplitude correlations in event-by-event fluctuating heavy-ion collisions}

\author{Jing Qian}
\email[Correspond to\ ]{qianjing8758@gmail.com}
\affiliation{Department of Physics, Harbin Institute of Technology, Harbin, 150001, People's Republic of China} 
\affiliation{Department of Physics, The Ohio State University, Columbus, OH 43210-1117, USA}
\author{Ulrich Heinz}
\affiliation{Department of Physics, The Ohio State University, Columbus, OH 43210-1117, USA}
\begin{abstract}
The effects of event-by-event fluctuations in the initial geometry of the colliding nuclei are important in the analysis of final flow observables in relativistic heavy-ion collisions. We use hydrodynamic simulations to study the amplitude correlations between different orders of event-by-event fluctuating anisotropic flow harmonics. While the general trends seen in the experimental data are qualitatively reproduced by the model, deviations in detail, in particular for peripheral collisions, point to the need for more elaborate future calculations with a hybrid approach that describes the late hadronic stage of the evolution microscopically. It is demonstrated explicitly that the observed anti-correlation between $v_2$ and $v_3$ is the consequence of approximately linear hydrodynamic response to a similar anti-correlation of the corresponding initial eccentricities $\epsilon_2$ and $\epsilon_3$. For $n{\,>\,}3$, the hydrodynamic correlations between $v_{2,3}$ and $v_n$ deviate from the rescaled correlations among the corresponding initial eccentricities, demonstrating nonlinear mode coupling effect in higher order flows.

\end{abstract}

\date{\today}

\maketitle

\section{Introduction}
\label{sec1}

Event-by-event (EbE) fluctuations of the density of produced matter in ultra-relativistic heavy-ion collisions arise from EbE fluctuations of the impacting nucleons' positions and those of the quark and gluon fields inside those nucleons. Anisotropic flow, which is generated by anisotropies in the pressure gradients, depends on the shape and structure of the initial density profile \cite{Miller:2003kd,Heinz:2013th}. The EbE fluctuations of the initial density profiles lead to the experimentally observed odd-order flow harmonics in symmetric collision systems \cite{Alver:2010gr} and to EbE fluctuations of and correlations among the flow coefficients and their corresponding flow angles \cite{Heinz:2013bua, Bhalerao:2011bp}. 

These flow fluctuations and correlations can be studied by using a variety of experimental observables  \cite{Bhalerao:2014xra}. Specific suggestions proposed during the past decade include the following: the distribution of $v_n$ and associated initial eccentricities $\epsilon_n$ \cite{Yan:2013laa, Yan:2014afa}, the correlation between different flow angles (event-plane correlators) \cite{Qiu:2012uy} and between the magnitudes of the flow coefficients \cite{Bilandzic:2013kga}, a principle component analysis (PCA) of flows \cite{Bhalerao:2014mua, Mazeliauskas:2015vea, Mazeliauskas:2015efa}, and the extraction of non-linear mode-coupling coefficients \cite{Yan:2015jma, Qian:2016fpi}. Several more can be found in the review~\cite{Jia:2014jca}.

Recently, the ATLAS and ALICE Collaborations performed measurements of correlations between the amplitudes of different anisotropic flows in Pb+Pb collisions at $\sqrt{s_{NN}} = 2.76$\,TeV at the LHC \cite{Jia:2014jca,Aad:2015lwa,Zhou:2015slf}. Using different methods, both groups report an anti-correlation between the elliptic and triangular flow coefficients, $v_2$ and $v_3$. The ALICE collaboration has studied symmetric 2-harmonic 4-particle cumulants to evaluate the correlation and compared them to transport and hydrodynamic simulations \cite{Zhou:2015slf}. The ATLAS collaboration found a {\it linear} anti-correlation between $v_3$ and $v_2$ but {\it non-linear} correlations between the quadrangular and pentangular flows $v_4,\ v_5$ and $v_2,\ v_3$, from which they extracted evidence for non-linear mode-coupling contributions 
to these higher-order flow harmonics \cite{Aad:2015lwa}. The ATLAS results have not yet been compared with dynamical evolution models.

In this paper, we study the correlations between different orders of flow using event-by-event hydrodynamic simulations with the (2+1)-dimensional code {\tt VISH2+1} \cite{Song:2007fn, Shen:2014vra}. Our goal is to see whether all aspects of the large set of correlation data reported by ATLAS in \cite{Aad:2015lwa} are in qualitative agreement with the hydrodynamic paradigm. Unfortunately, exploring detailed correlations among the event-by-event fluctuating flow coefficients requires large statistics not only on the experimental side, but also theoretically, where large numbers of events with fluctuating initial conditions must be simulated dynamically. To keep the numerical effort manageable we will perform the simulations in pure hydrodynamic mode, i.e. without switching to a microscopic description of the late hadronic stage. While such a hybrid approach will eventually be required for a fully quantitative comparison with the experimental data, the present study should be sufficient to recognize serious discrepancies with the hydrodynamic approach that might threaten to invalidate it. In this study we do not systematically explore the sensitivity of these correlations to the QGP shear viscosity $\eta/s$ and rather fix it to the value 0.08 which, for the Monte Carlo Glauber initial conditions used here, is preferred by the $p_T$ distributions of identified hadrons and of the charged hadron elliptic and triangular flow \cite{Song:2010mg}.\footnote{%
     \label{fn1}%
     \baselineskip=12pt%
     In Appendix~\ref{appb} we use a previously generated smaller set of hydrodynamic events with
     Monte Carlo KLN initial conditions \cite{Qian:2016fpi}, which were evolved with three different values
     of the specific shear viscosity $\eta/s\eq0,\,0.08$, and 0.2, to discuss different possible shapes of the 
     ``boomerang''-like dependence of $v_m$ on $v_n$ for different collision centralities that was 
     pointed out by the ATLAS Collaboration \cite{Aad:2015lwa}, and how these shapes vary with $\eta/s$.
     }
Since the flow fluctuations obtained from our pure hydrodynamic simulations are not expected to be quantitatively precise, we will refrain here from a direct comparison with the experimental data, but instead invite the reader to compare with the ATLAS data \cite{Aad:2015lwa}.

The hydrodynamic $v_m-v_n$ correlations are presented in Sec.~\ref{sec2}. In Sec.~\ref{sec3}, we discuss the linear and nonlinear contributions to higher-order flows and to their correlations with $v_2$ and $v_3$. Our results are discussed and summarized in Sec.~\ref{sec4}. Details of the event-shape selection used in our analysis are presented in Appendix~\ref{appa} while the relationship between the boomerang-like shape of the $v_n$-$v_m$ correlations and the centrality dependence of the harmonic flow coefficients is discussed in Appendix~\ref{appb}.

\section{$\boldmath{v_m- v_n}$ correlations}
\label{sec2}

The anisotropic flow coefficients $v_n$ are defined as the $n$th azimuthal Fourier coefficient of the momentum spectrum. It has been pointed out \cite{Kolb:2003zi, Borghini:2005kd} that in the absence of event-by-event fluctuation (i.e. for hydrodynamics with smooth initial condition) even flow harmonics are  correlated because, even if the fluid velocity profile is only elliptically deformed, a full set of even flow harmonics is in general generated because the fluid velocity enters through the exponent of the (flow-boosted) thermal distribution on the freeze-out surface. When EbE density fluctuations in the initial state are included, the resulting flow fluctuations of different harmonic orders (both even and odd) are, in general correlated by geometric constraints on the positions and shapes of these fluctuations inside the spatially deformed region where the colliding nuclei overlap. For example, by selecting events in which two suitably located upward fluctuations of the density generate an especially large geometric ellipticity, this selection restricts the possibility for adding a third hot spot to generate also large triangularity of the density distribution. And if such high-ellipticity events created by hot spots do feature also a triangular deformation, the axes of the corresponding ellipses and triangles (i.e. the elliptic and triangular participant planes) tend to be correlated \cite{Nagle:2010zk, Huo:2013qma}. Hydrodynamics will translate these correlations between the harmonic eccentricity vectors in the initial state into corresponding correlations among the final harmonic flow coefficients.

In this section, we explore these final-state flow correlations with the (2+1)-dimensional viscous hydrodynamic model {\tt VISH2+1}, using MC-Glauber initial condition and minimal specific shear viscosity $\eta/s = 0.08$. We use the same setup as in a previous work \cite{Qian:2016fpi}, starting the hydrodynamic evolution at longitudinal proper time $\tau_0\eq0.6$\,fm/$c$ and ending it on an isothermal decoupling surface of temperature $T_\mathrm{dec}\eq120$\,MeV. Guided by the experimental analysis in \cite{Aad:2015lwa} we defined 14 equal centrality bins (0\%-5\%, 5\%-10\%, \dots, 65\%-70\%, based on their final charged multiplicity density at midrapidity) and generated 3000 hydrodynamically evolved events in each bin. For each of the multiplicity bins these 3000 events were then ordered and binned by their $q_2$ or $q_3$ values, where $\bm{q}_n\eq{q}_n\,e^{in\Psi^q_n}\eq\langle m_T e^{in\phi_p}\rangle/\langle m_T\rangle$ was calculated from the Cooper-Frye spectrum on the freeze-out surface as the transverse energy ($m_T\eq\sqrt{m^2{+}p_T^2}$) weighted average of the phase factors $e^{in\phi_p}$ \cite{Aad:2015lwa}\footnote{%
     \label{fn2}%
     \baselineskip=12pt%
     In the experimental analysis \cite{Aad:2015lwa} $q_n$ was measured at forward rapidity, 
     $3.3{\,<\,}|\eta|{\,<\,}4.8$, which leads to some decorrelation from the midrapidity anisotropic flows 
     $v_n$ caused by rapidity-dependent fluctuations of the multiplicity density and initial transverse density 
     distribution \cite{Huo:2013qma, Pang:2015zrq, Broniowski:2015oif, Khachatryan:2015oea, %
     Aaboud:2016jnr} that we cannot simulate with our longitudinally boost-invariant evolution code. 
     We therefore compute $q_n$ at midrapidity.
     } 
(whose unweighted average over the spectrum defines the anisotropic flow coefficients $V_n\eq{v}_n\,e^{in\Psi_n}$). (Binning events by a certain event characteristic such as $q_n$ that can be measured event-by-event is known as ``event-shape engineering'' \cite{Schukraft:2012ah}.) 
 
%
\begin{figure*}
    \vspace*{-8mm}
    \includegraphics[width=\textwidth,height=8cm]{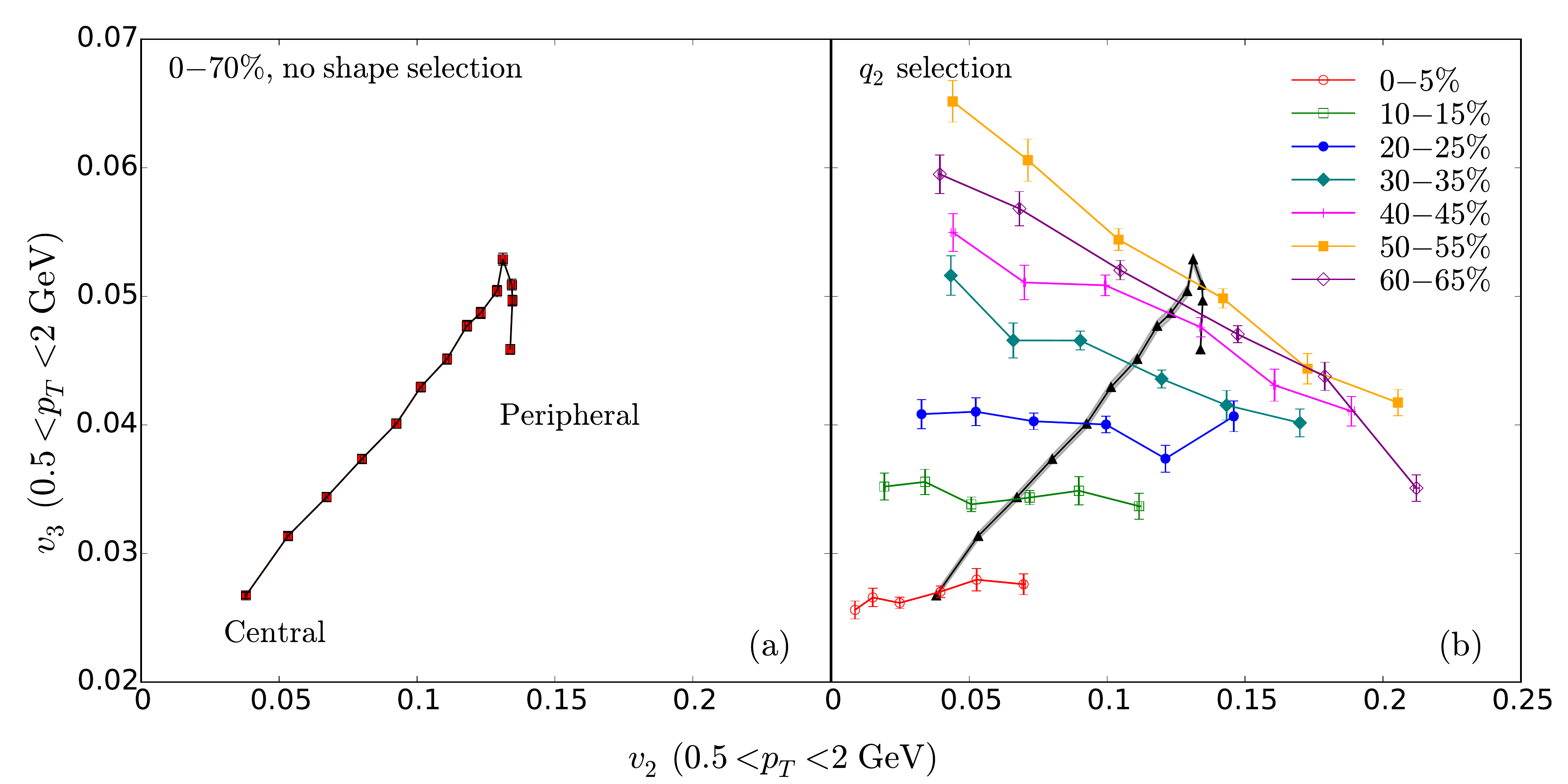}\\
    \includegraphics[width=\textwidth,height=8cm]{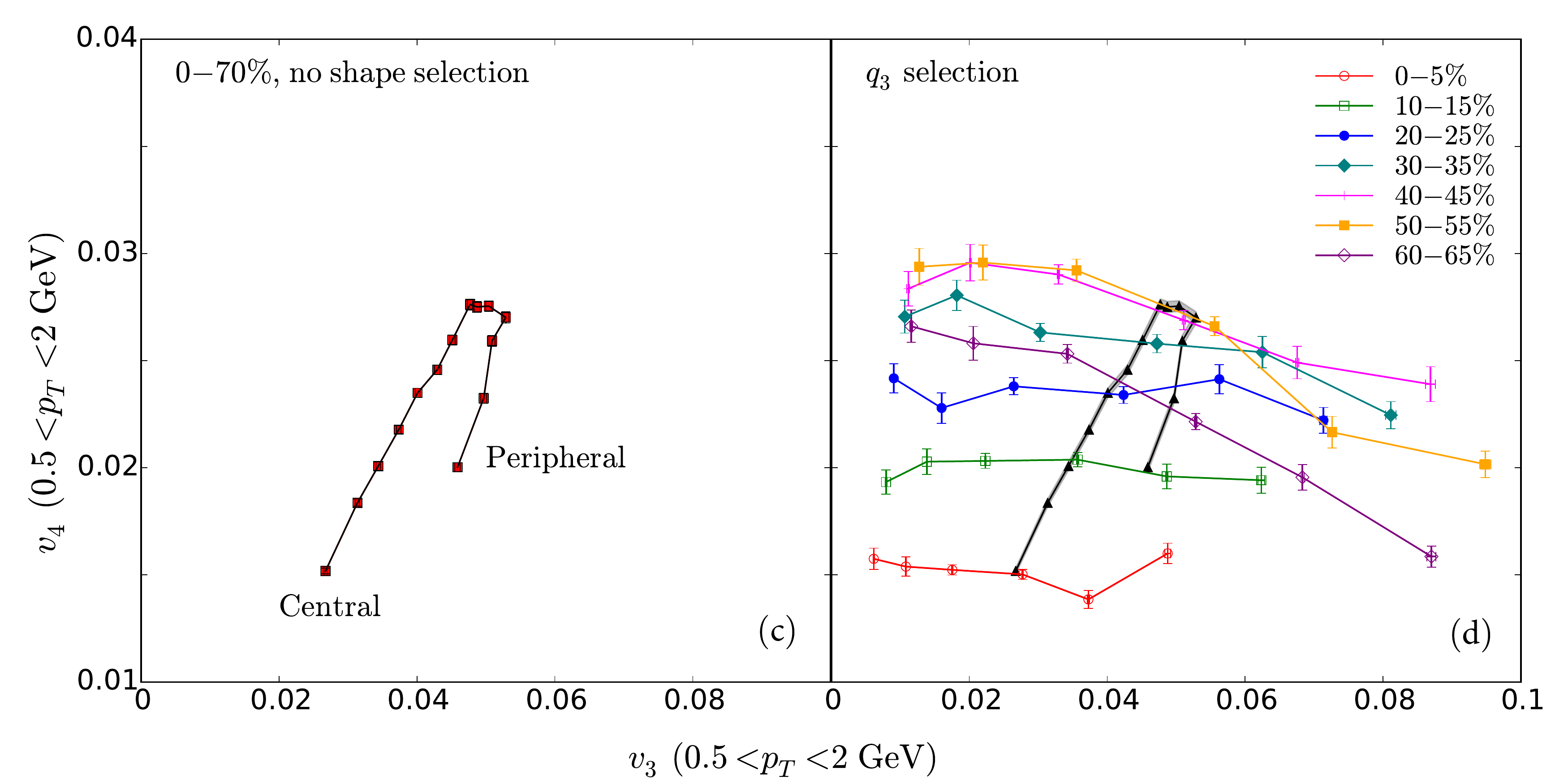}
    \vspace*{-8mm}
    \caption{(Color online)
    \baselineskip=14pt
    The correlation between $v_3$ and $v_2$ (a,b) and between $v_4$ and $v_3$ (c,d) as a function 
    of event centrality for the 14 centrality bins described in the text, without event-shape selection (a,c) 
    and for events with different $q_2$ values (b) or $q_3$ values (d), sorted by event shape into 6 $q$ 
    bins as described in the text. The results were obtained from the viscous hydrodynamic simulations 
    of Pb-Pb collisions at the LHC described in the text and should be compared with the experimental 
    correlations obtained by the ATLAS Collaboration and shown in Figs.~7 and 12 of 
    Ref.~\cite{Aad:2015lwa}. Note that what is plotted is the rms value of $v_n$ ($v_n\{2\}$) for charged
    hadrons including resonance decay contribution, integrated over the indicated $p_T$ range; the 
    qualifier $\{2\}$ is dropped for clarity.
    \vspace*{-3mm}
     \label{F1}
}
\end{figure*}
%

Figure~\ref{F1}a plots the rms triangular flow $v_3\{2\}$ against the rms elliptic flow $v_2\{2\}$ of charged hadrons for the 14 centrality bins defined above; Fig.~\ref{F1}c shows the same for $v_4$ vs. $v_3$.  Qualitatively, the two plots exhibits the boomerang-like relation between the rms triangular and elliptic flow coefficients as a function of collision centrality that was first pointed out by the ATLAS Collaboration \cite{Aad:2015lwa}. However, the detailed shape of the boomerang predicted by our viscous hydrodynamic calculations disagrees with the one measured experimentally (see Figs.~7 and 12 in Ref.~\cite{Aad:2015lwa}). In Appendix~\ref{appb} we discuss how the shape of the ``boomerang'' relates to the centrality dependences of the two flow harmonics plotted against each other in the graph and demonstrate that this shape can change qualitatively (including ``looping boomerangs'') for different values of the QGP shear viscosity $\eta/s$ and for different combinations of harmonics $m$ and $n$. The failure of the pure hydrodynamic results shown in Fig.~\ref{F1}a to correctly reproduce the experimentally measured $v_3-v_2$ correlation can, at least in part, be attributed to an overprediction of both $v_2$ and $v_3$ in peripheral collisions; it is possible that modeling the evolution with a hybrid code that accounts for the larger effective shear viscosity of the evolving matter in its less strongly coupled late hadronic stage may correct this in the future.

Figures~\ref{F1}b,d show what happens when we subdivide the 3000 events from each centrality bin into 6 event-shape selected event classes with different $q_2$ resp. $q_3$ values \cite{Schukraft:2012ah, Petersen:2013vca, Huo:2013qma}. To do so we order the events by $q_2$ and throw them into bins covering the following fractions of these ordered events: $0{-}0.1,\, 0.1{-}0.2,\, 0.2{-}0.5,\, 0.5{-}0.8,\, 0.8{-}0.9,$ and $0.9{-}1.0$. The choice of the $q_n$ bins is discussed in more detail in Appendix ~\ref{appa}. The hydrodynamic model predictions shown in Fig.~\ref{F1}b,d should be compared to Figs.~7b and 12b in Ref.~\cite{Aad:2015lwa}. Fig.~\ref{F1}b qualitatively confirms the anti-correlation between $v_3$ and $v_2$ within fixed multiplicity bins observed by ATLAS \cite{Aad:2015lwa}. However, while in the experimental data this negative correlation persists also (albeit weakly) in the most central collisions, the hydrodynamic model simulations predict a weakly positive correlation between $v_2$ and $v_3$ in central collisions. As will be further discussed below in Fig.~\ref{F2}, this positive correlation tracks a similar positive correlation between the corresponding initial eccentricities $\epsilon_2$ and $\epsilon_3$ in the MC-Glauber model. This is not unexpected because it is well known that the hydrodynamic response of $v_2$ and $v_3$ to $\epsilon_2$ and $\epsilon_3$, respectively, is to very good approximation linear, especially at small eccentricities.\footnote{%
     \label{fn3}%
     \baselineskip=12pt%
     We found that the weak positive correlation between $\epsilon_2$ and $\epsilon_3$ in central 
     Pb+Pb collisions is an artifact arising from the neglect of p-p multiplicity fluctuations in the version 
     of the MC-Glauber model used in this work and also in \cite{Aad:2015lwa}. This unfortunate choice 
     was motivated by our desire to re-use a significant number of previously generated hydrodynamic 
     events for this analysis and to avoid the retuning of initial and freeze-out parameters of the 
     hydrodynamic module to a new initial-state model. Without p-p multiplicity fluctuations, both 
     $\epsilon_2$ and $\epsilon_3$ increase strongly with multiplicity within the $0-5\%$ centrality bin,
     and the positive correlation between them seen in Figs.~\ref{F1}b, \ref{F2}a for that bin reflects 
     mostly this centrality dependence of $\epsilon_{2,3}$ instead of genuine shape change at fixed 
     centrality \cite{private}. When we add p-p multiplicity fluctuations in the MC-Glauber model as 
     described in \cite{Shen:2014vra}, the slight positive correlation between $\epsilon_3$ and 
     $\epsilon_2$ seen in Fig.~\ref{F2}a below for $0-5\%$ centrality disappears and even turns into 
     a slight anti-correlation (albeit a very weak one, weaker than the one seen in Fig.~\ref{F2}b). 
     Linear hydrodynamic response will turn this into a weak anti-correlation between $v_3$ and $v_2$ 
     at $0-5\%$ centrality, similar to what was observed by ATLAS \cite{Aad:2015lwa}. Due to the 
     numerical expense we have, however, not rerun the modified initial conditions through 
     hydrodynamics. 
     }

Figure~\ref{F1}d shows a weak anti-correlation between $v_4$ and $v_3$ within events of fixed multiplicity (centrality). While this is in qualitative agreement with the ATLAS data shown in Fig.~12 of Ref.~\cite{Aad:2015lwa}, our simulations predict that this anti-correlation strengthens in more peripheral collisions -- a feature that is not obvious in the ATLAS data. Again, we note that\break 
\newpage\noindent
our pure hydrodynamic simulations overpredict the mean and variance of the $v_3$ distributions in the more peripheral centrality classes compared to the ATLAS data.   

%
\begin{sidewaysfigure}
    \includegraphics[width=\textwidth,angle=0]{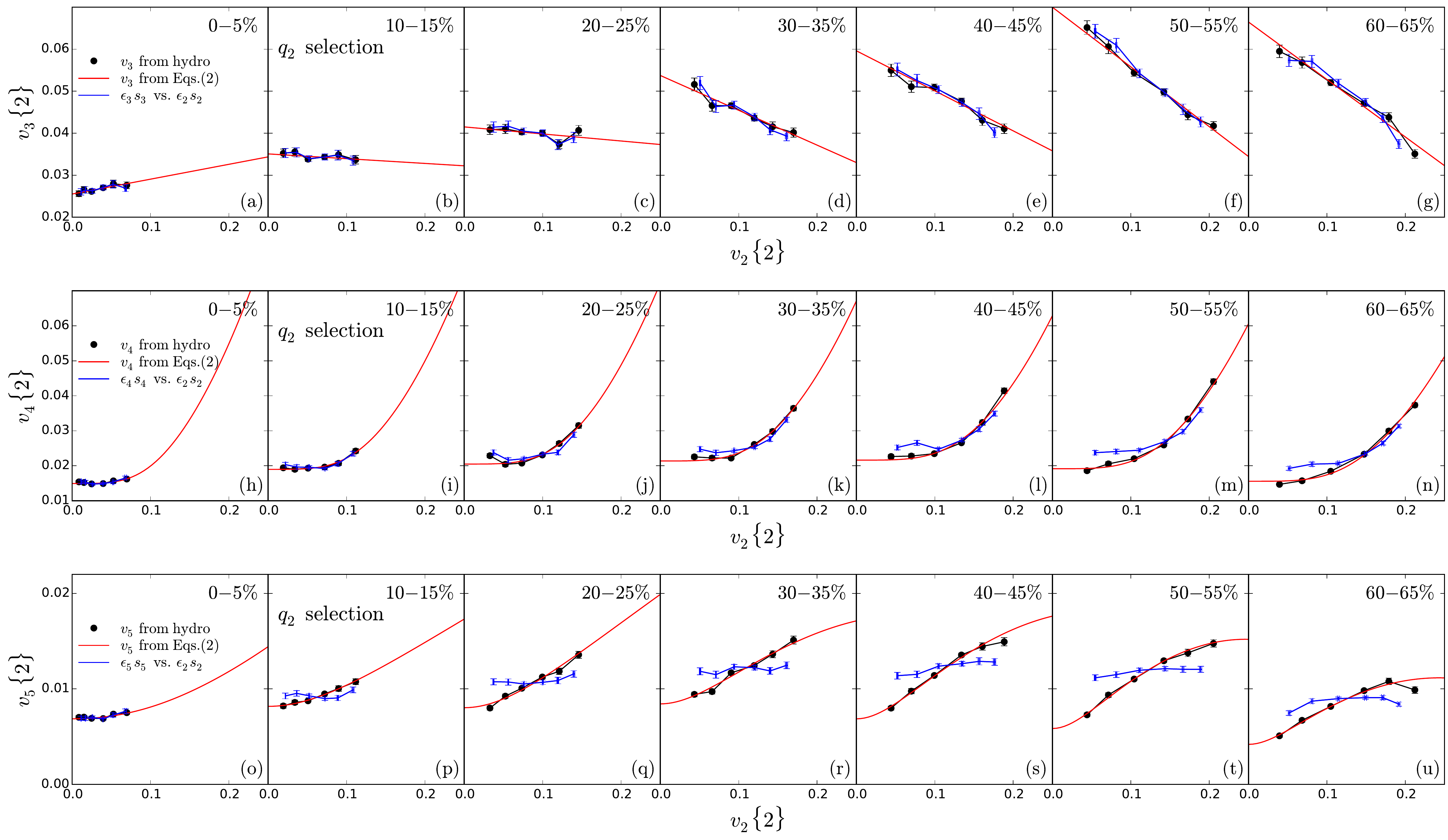}
    \caption{(Color online)
    \baselineskip=13pt
    The $v_n$-$v_2$ correlations for $n\eq3,4,5$, for six $q_n$ selections in seven centrality 
    intervals of 5\% width (only every other centrality bin is shown), using viscous hydrodynamics 
    with MC-Glauber initial condition, $\eta/s=0.08$.  Black circles show the hydrodynamic 
    flow amplitude correlations, blue lines are the corresponding eccentricity correlations, with eccentricities 
    $\epsilon_n$ rescaled according to Eq.~(\ref{en_vn}). The red lines show fits according to 
    Eqs.~(\ref{fit}). Error bars indicate the statistical uncertainties arising from the finite number of 
    events in each $q$-bin. For easier comparison with the ATLAS data \cite{Aad:2015lwa} the 
    charged hadron flows are integrated over the same $p_T$ range as in the experiment.
     \label{F2}
}
\end{sidewaysfigure}
%

%
\begin{sidewaysfigure}
    \includegraphics[width=\textwidth,angle=0]{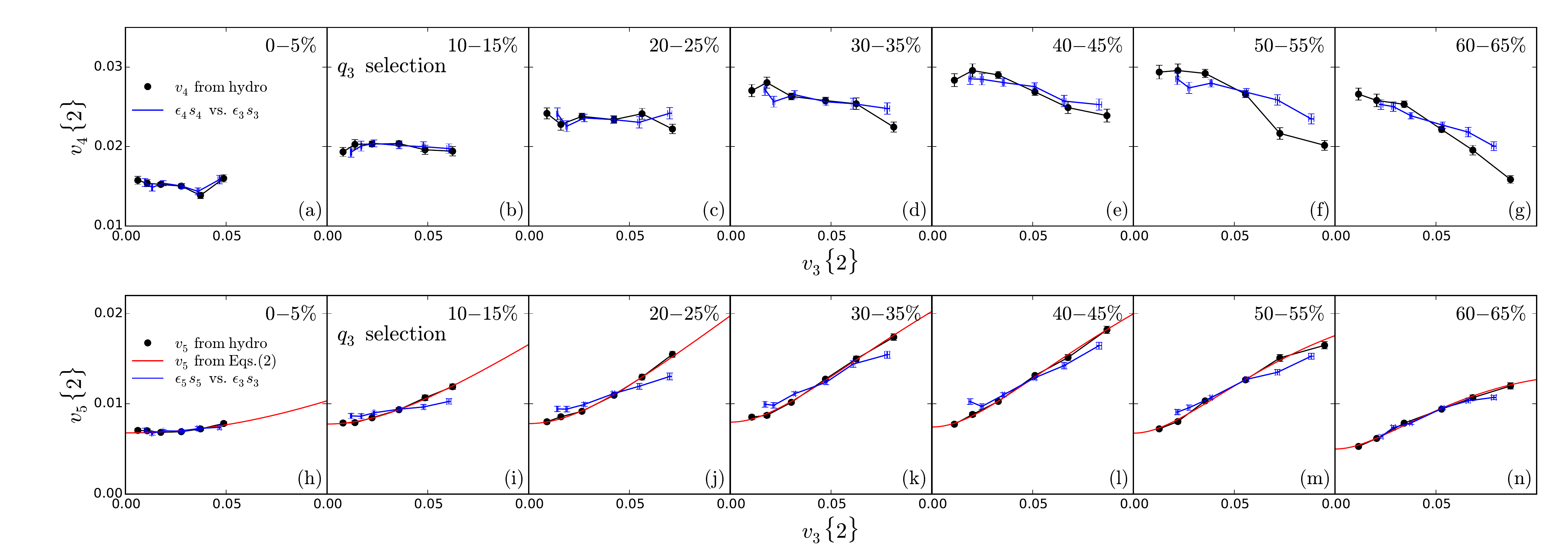}    
    \caption{(Color online)
    Similar to Fig.~\ref{F2}, but for $v_n$-$v_3$ correlations ($n\eq4,5$).
    \label{F3}
}
\end{sidewaysfigure}
%

In Figs.~\ref{F2} and \ref{F3} we compare, for every other centrality bin, the correlations between $v_{3,4,5}$ and $v_2$ (Fig.~\ref{F2}) and between $v_{4,5}$ and $v_3$ (Fig.~\ref{F3}) with the correlations between the corresponding initial eccentricities $\epsilon_n$, using event-shape selected events ordered by their $q_2$ and $q_3$ values, respectively. For this comparison the initial eccentricities in each $q$-bin were rescaled by the ratios between the rms flows and eccentricities in the given centrality bin as suggested in \cite{Aad:2015lwa}; that is, we compare the hydrodynamically simulated $v_m-v_n$ correlations (black circles connected by black lines) with corresponding $\tilde{v}_m-\tilde{v}_n$ correlations (blue lines) where $\tilde{v}_n$ is calculated in each $q$-bin as 
\begin{equation}
  \tilde{v}_n\{2\}(q_n)  \equiv s_n\,\epsilon_n\{2\}(q_n) 
\label{en_vn}
\end{equation}
with a scaling factor $s_n = \sqrt{\langle v_n^2 \rangle / \langle \epsilon_n^2 \rangle}=v_n\{2\}/\epsilon_n\{2\}$ determined from the ensemble average over the entire centrality bin. We see in the top row of Fig.~\ref{F1} (panels a-g) that the (anti-)correlation between $v_3$ and $v_2$ is linear and, except for small deviations in the outer $q_2$ bins (i.e. in the tails of the $q_2$-distribution), tracks the corresponding eccentricities. For all other combinations of $n$ and $m$ (see bottom two rows of Fig.~\ref{F2} and Fig.~\ref{F3}), we observe a similar agreement between the $v_m-v_n$ correlations and the corresponding scaled eccentricity correlations only for the most central ($0-5\%$ centrality) collisions. In such collisions, there is no geometric contribution to the eccentricities, i.e. all $\epsilon_n$ are entirely due to initial-state fluctuations. For all other centrality bins, the even eccentricity harmonics, in particular the ellipticity $\epsilon_2$, has a nonzero geometric component due to the almond shape of the nuclear overlap region. As also observed in the experimental data \cite{Aad:2015lwa}, this leads to increasingly strong mode-mixing effects at larger impact parameters that lead to deviations between the $v_m-v_n$ correlations and those between the corresponding scaled eccentricities $\tilde{v}_m-\tilde{v}_n$ for the more peripheral centrality classes.\footnote{%
    \label{fn4}%
    \baselineskip=12pt%
    Note that the correlations between the scaled eccentricities $\tilde{v}_m$ and $\tilde{v}_n$ 
    for $(m,n)\ne(3,2)$ are also in general non-linear; if the hydrodynamic response of $v_n$ to 
    $\epsilon_n$ were linear for all values of $n$, however, we would expect the $v_m-v_n$ correlations
    to perfectly track the corresponding $\tilde{v}_m-\tilde{v}_n$ correlations, whatever their shape. The 
    fact that they don't indicates non-linear mode-mixing in the hydrodynamic response.
    }
Quite generically we see from Fig.~\ref{F2} that $v_4$ and $v_5$ increase with $v_2$ more strongly than $\epsilon_4$ and $\epsilon_5$ increase with $\epsilon_2$, indicating a non-linear mode-coupling contribution from $\epsilon_2$ (which increases with impact parameter). Similarly, Fig.~\ref{F3} shows that $v_5$ increase with $v_3$ more strongly than $\epsilon_5$ increase with $\epsilon_3$, indicating a similar non-linear mode-coupling contribution from $\epsilon_3$ (which, according to Fig.~\ref{F1}a, increases with impact parameter up to $50\%-55\%$ centrality). On the other hand, no such clear nonlinear contribution is seen for $v_4$ as a function of $v_3$ (top row in Fig.~\ref{F3}), indicating the absence of appreciable nonlinear mode coupling between these two harmonics.

To quantify the mode-coupling effects seen in Figs.~\ref{F2}, \ref{F3} we fit the hydrodynamic flow coefficients in each centrality bin to the following functional forms \cite{Aad:2015lwa}:
\begin{eqnarray}
\label{fit}
  && v_3\{2\} = v_3^0 + k_3\, v_2\{2\},
  \nonumber\\
  && v_4\{2\} = \sqrt{(v_4^0)^2 + (k_4\, v_2^2\{2\})^2}, 
  \\\nonumber
  && v_5\{2\} = \sqrt{(v_5^0)^2 + (k_5\, v_2\{2\}\, v_3\{2\})^2}.
\end{eqnarray}
The corresponding fits are shown in the figures as red solid lines.\footnote{%
    \label{fn5}%
    \baselineskip=12pt%
    Note that the fit of $v_5$ according to Eq.~(\ref{fit}) depends on whether $q_2$ or $q_3$ is used 
    for event-shape selection. To fit the $v_5\{2\}-v_2\{2\}$ correlation within a given centrality class, 
    we use the ansatz $v_3\{2\} = v_3^0 + k_3\, v_2\{2\}$ in the fit function for $v_5\{2\}$, 
    $v_5\{2\} = \sqrt{(v_5^0)^2 + [k_5\, v_2\{2\}\, (v_3^0{+}k_3\, v_2\{2\})]^2}$, with $v_3^0$ and $k_3$ 
    obtained from fitting the $v_3-v_2$ correlation at the same centrality using $q_2$ event-shape 
    selection. For the fit of the $v_5\{2\}-v_3\{2\}$ correlation at the same centrality we similarly use 
    the fit function $v_5\{2\} = \sqrt{(v_5^0)^2 + [\frac{k_5}{k_3}\, v_3\{2\}\, (v_3\{2\}{-}v_3^0)]^2}$ but 
    obtain $v_3^0$ and $k_3$ from a fit of the $v_2-v_3$ correlation at the same centrality after 
    binning the events in $q_3$.
}    
From the discussion above it is clear that the mode-coupling coefficients $k_{3,4,5}$ encode a combination of the initial correlations among the corresponding eccentricity coefficients and hydrodynamic mode-mixing during the subsequent dynamical evolution. A discussion of the centrality dependence of the fit parameters $k_n$ and $v_n^0$ is presented in the next section.

Before closing this section let us briefly comment on the weak anti-correlation between $v_4$ and $v_3$ seen in the top row of Fig.~\ref{F3}. This anti-correlation strengthens as the collisions become more peripheral. At the same time, deviations from a qualitatively similar but weaker anti-correlation between the associated (scaled) initial eccentricities $\epsilon_4$ and $\epsilon_3$ also become larger. We believe that this is caused by the well-known nonlinear mode-mixing contribution to $v_4$ from $v_2^2$ (or $\epsilon_2^2$) which increases with impact parameter and (according to Fig.~\ref{F1}b) is anti-correlated with $v_3$.
 
\section{Linear/nonlinear decomposition of $v_4$ and $v_5$}
\label{sec3}
%
\begin{figure*}[!hbt]
    \includegraphics[width=\linewidth]{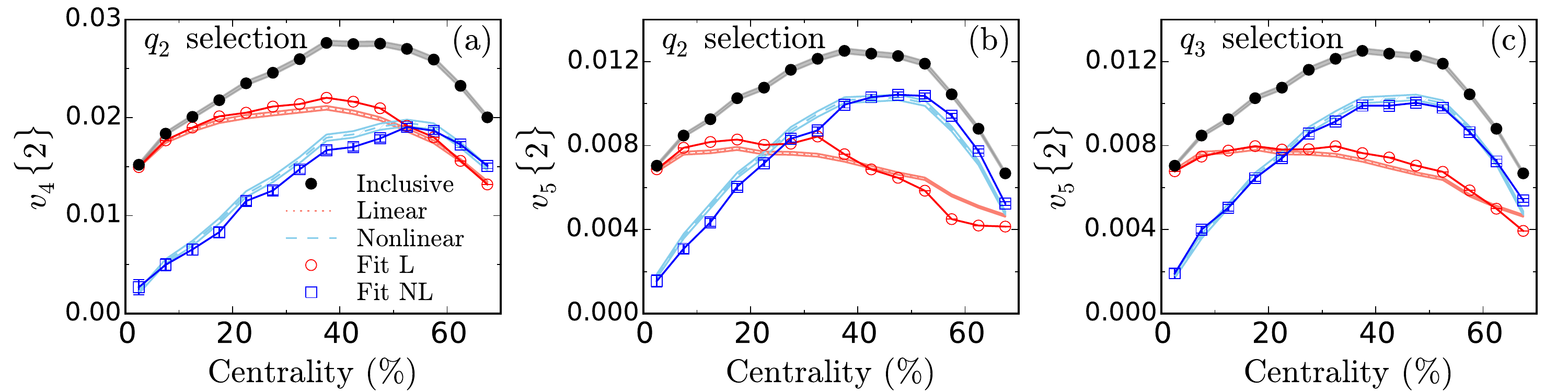}
    \caption{(Color online)
    \baselineskip=14pt
    Centrality dependence of $v_4$ (a) and $v_5$ (b,c), together with their associated linear and 
    nonlinear components defined in Eqs.~(\ref{v4v5}) and (\ref{v4v5_fit}), from the hydrodynamic 
    simulations described in the text. Black solid circles show the full $v_n\{2\}$. Red dotted and blue 
    dashed lines are the linear (L) and nonlinear (NL) parts as defined in Eqs.~(\ref{v4v5}) and are 
    identical in panels b and c. Red circles (``Fit L'') and blues squares (``Fit NL'') show the linear 
    and nonlinear components $v_n^\mathrm{L,\, fit}\{2\}$ and $v_n^\mathrm{NL,\, fit}\{2\}$, 
    respectively, defined in Eqs.~(\ref{v4v5_fit}) based on the fit functions (\ref{fit}); they differ between
    panels b and c. The shaded regions show the statistical uncertainties associated with the finite 
    number (3000) of collision events per centrality bin.
     \label{F4}
}
\end{figure*}
%

Through hydrodynamic evolution, anisotropic pressure gradients build up anisotropic particle distributions in transverse momentum space, i.e. anisotropic flow. Hydrodynamic calculations show that $v_2$ and $v_3$ respond approximately linearly to the corresponding initial\break
\newpage\noindent
eccentricities, e.g. $v_n/\epsilon_n = \mathrm{const}$, except for large impact parameters \cite{Qiu:2011iv, Gardim:2011xv, Niemi:2012aj, Fu:2015wba, Niemi:2015qia, Noronha-Hostler:2015dbi}. Higher-order flow harmonics, on the other hand, exhibit non-linear response, i.e. mode-mixing \cite{Qiu:2011iv, Gardim:2011xv, Teaney:2012ke}.

The authors of \cite{Yan:2015jma} suggested that the higher-order complex harmonic flow coefficients $V_n \,(n>3)$ could be modeled as sums of linear and nonlinear response terms: $V_n = V_n^\mathrm{L} + V_n^\mathrm{NL}$. Taking $V_n^\mathrm{L}$ and $V_n^\mathrm{NL}$ are uncorrelated \cite{Yan:2015jma, Qian:2016fpi} and $\langle V_n^\mathrm{L} \rangle = 0$, $v_n\{2\}$ can thus be decomposed as follows:
\begin{eqnarray}
\label{vn_decomposition}
  v_n^2\{2\} &=& \langle |V_n|^2 \rangle
                    = \langle |V_n^\mathrm{L} + V_n^\mathrm{NL}|^2 \rangle
  \nonumber\\
                   &=& \langle |V_n^\mathrm{L}|^2 \rangle + \langle |V_n^\mathrm{NL}|^2 \rangle 
                        + 2 \mathrm{Re} (\langle V_n^\mathrm{L} \rangle \langle V_n^\mathrm{NL*} \rangle)
   \nonumber\\
                   &=& (v_n^\mathrm{L}\{2\})^2 + (v_n^\mathrm{NL}\{2\})^2.
\end{eqnarray}
In other words, the mean square of the elliptic flow is equal to the sum of the mean squares of its linear and nonlinear parts. Further, the decomposition of $V_4$ and $V_5$ suggested in \cite{Yan:2015jma},
\begin{eqnarray}
  &&V_4  = V_{4 \mathrm{L}} + \chi_{422} V_2^2,
  \nonumber\\
  &&V_5  = V_{5 \mathrm{L}} + \chi_{523} V_2 V_3,
\label{VlVnl}
\end{eqnarray}
allows $v_n^\mathrm{L}\{2\}$ and $v_n^\mathrm{NL}\{2\}$ for $n=4,5$ to be expressed as \cite{Yan:2015jma, Qian:2016fpi}
\begin{eqnarray}
\label{v4v5}
v_4^\mathrm{L} &=& \sqrt{\langle v_4^2 \rangle - \frac{|\mathrm{Re} \langle V_4 V_2^{*2} \rangle|^2}{\langle v_2^4 \rangle}},\quad 
v_4^\mathrm{NL} = \frac{|\mathrm{Re} \langle V_4 V_2^{*2} \rangle|}{\sqrt{\langle v_2^4 \rangle}},
\nonumber\\
v_5^\mathrm{L} &=& \sqrt{\langle v_5^2 \rangle - \frac{|\mathrm{Re} \langle V_5 V_2^* V_3^* \rangle|^2}{\langle v_2^2 v_3^2 \rangle}},\quad 
v_5^\mathrm{NL} = \frac{|\mathrm{Re} \langle V_5 V_2^* V_3^* \rangle|}{\sqrt{\langle v_2^2 v_3^2 \rangle}}.
\end{eqnarray}
(Here we suppressed for clarity the label $\{2\}$ denoting the second order cumulant.) We note that this definition differs from a alternative estimate used by ATLAS (see Eqs.~(18) and (21) in \cite{Aad:2015lwa}) which is based on event plane correlators and includes additionally nonlinear mode coupling effects among the flow angles.

The fit functions (\ref{fit}), on the other hand, suggest the following decomposition of $v_4$ and $v_5$ into linear and nonlinear parts (also used in \cite{Aad:2015lwa}):
\begin{eqnarray}
\label{v4v5_fit}
v_4^\mathrm{L,\, fit} &=& v_4^0,\quad 
v_4^\mathrm{NL,\, fit} = \sqrt{v_4\{2\}^2 - (v_4^0)^2} = k_4\, v_2\{2\}^2.\\
v_5^\mathrm{L,\, fit} &=& v_5^0,\quad 
v_5^\mathrm{NL,\, fit} = \sqrt{v_5\{2\}^2 - (v_5^0)^2} = k_5\, v_2\{2\}\, v_3\{2\}.
\nonumber
\end{eqnarray}

The different prescriptions (\ref{v4v5}) and (\ref{v4v5_fit}) for separating $v_n$ into linear and nonlinear contributions are compared for $n\eq4$ and 5 (i.e. for the quadrangular and pentangular flows) in Fig.~\ref{F4} as a function of collision centrality.\footnote{%
    \label{fn6}%
    \baselineskip 12 pt%
    These figures are to be compared with Figs.~11 and 15 in Ref.~\cite{Aad:2015lwa} where the 
    number of participating nucleons $N_\mathrm{part}$ from the Glauber model is used as a 
    measure for collision centrality.
    }
We see that the two prescriptions yield compatible results: In central collisions, the linear contribution dominates the total flow, whereas the nonlinear contribution increases with increasing impact parameter (at least in part through mode-mixing effects involving $\epsilon_2$, as discussed in the preceding section). While the centrality dependence of the linear contributions to $v_4$ and $v_5$ is weak, the nonlinear contribution varies strongly with collision centrality. It is noteworthy that the separation of $v_5$ into linear and non-linear parts according to Eqs.~(\ref{fit},\ref{v4v5_fit}) shows only very weak sensitivity to whether the fit is based on event-shape selection using $q_2$ or $q_3$. This is discussed in more detail in Appendix~\ref{appa}. 

\section{Summary and further discussion}
\label{sec4}

In a beautiful experimental analysis \cite{Aad:2015lwa}, the ATLAS Collaboration recently performed a comprehensive analysis of correlations among the fluctuating anisotropic flow amplitudes $v_n$ in Pb+Pb collisions at the LHC, using  a set of correlators that are completely insensitive to the associated fluctuating flow angles $\Psi_n$. The purpose of the present work was to study whether the experimentally measured flow amplitude correlations and their dependence on collision centrality can be described and understood within the otherwise highly successful hydrodynamic approach to the dynamical evolution of relativistic heavy ion collisions. 

We found qualitative agreement with all the main characteristics of the experimentally observed flow correlations. Quantitatively the hydrodynamic model simulations fail in the most peripheral collisions where all anisotropic flow coefficients are overpredicted. We ascribe this failure predominantly to the lack of a proper microscopic description of the late hadronic stage in our pure hydrodynamic simulations which did not include a transition from fluid dynamics to a hadron cascade below the hadronization temperature. This is known to underestimate shear viscous effects and the associated suppression of anisotropic flow build-up in the hadronic phase. We showed that the correlations among the anisotropic flow amplitudes $v_n$ in part reflect similar correlations among the corresponding eccentricities $\epsilon_n$ of the fluctuating density profiles in the initial state, enhanced, however, by nonlinear mode-coupling effects in the hydrodynamic evolution. For elliptic and triangular flow we found a linear anti-correlation that reflects a linear hydrodynamic mapping of a similar linear anti-correlation between the initial elliptic and triangular eccentricities. For higher-order flow harmonics the correlations between the corresponding initial eccentricities are found in general to be non-linear; while their hydrodynamic mapping to flow correlations is found to be linear in central collisions, it becomes increasingly nonlinear in non-central collisions due to mode-coupling effects involving the growing elliptic flow $v_2$. We decomposed the $v_m-v_n$ correlations into linear and non-linear contributions whose centrality dependence qualitatively agrees with the experimental data; two different procedures for this decomposition yielded mutually compatible results. While the linear component depends only weakly on the collision centrality, the nonlinear component varies strongly with impact parameter. We emphasize, however, that the non-linear part of the correlation extracted by these methods includes nonlinearities from both the initial state (through nonlinear eccentricity correlations) and from the hydrodynamic evolution (through mode-coupling effects) which cannot be separated model-independently. It is likely that both types of nonlinearities, the ones in the initial state and the hydrodynamic ones, are caused primarily by the elliptic geometric deformation of the nuclear overlap region in noncentral Pb+Pb collisions. This conjecture can be tested experimentally and theoretically in ultra-central U+U collisions where the nuclear overlap region is elliptically deformed even at zero impact parameter.

In Appendix~\ref{appb} we showed that the detailed shape of the centrality dependence of hydrodynamic $v_m-v_n$ correlations is quite sensitive to the shear viscosity of the expanding fluid, and also to the fluctuation spectrum of the initial state of the expanding fireball. This offers the hope that such correlations can in future studies be used as valuable constraints for both initial state fluctuation and QGP transport coefficients. This will, however, require full simulations with a hybrid model approach that correctly implements the early and late non-equilibrium dynamics of the evolving nuclear fireball.

\begin{acknowledgments}
The authors thank Jiangyong Jia, Jia Liu, Christopher Plumberg and Chun Shen for discussions. Computing resources were generously provided by the Ohio Supercomputer Center \cite{OhioSupercomputerCenter1987}. This work was supported by the U.S.\ Department of Energy, Office of Science, Office for Nuclear Physics Office under Award \rm{DE-SC0004286} and, in part, through the Beam Energy Scan Theory (BEST) Topical Collaboration.
\end{acknowledgments}

\appendix
\vspace*{-3mm}
\section{$v_m-v_n$ structure}
\label{appb}

In this Appendix we explore the various possible types of $v_m-v_n$ correlation structures (``boomerang shapes'', such as the one shown in  Fig.~\ref{F1}a) that can arise from qualitative differences in the centrality dependences of the two flow harmonics $v_n$, $v_m$ that are being correlated. No event-shape selection is performed in this Appendix, the events are only sorted by multiplicity (more precisely, by the initial entropy $dS/dy$ which is monotonically related to the final multiplicity). 

%
\begin{figure*}[!hbt]
    \includegraphics[width=\linewidth]{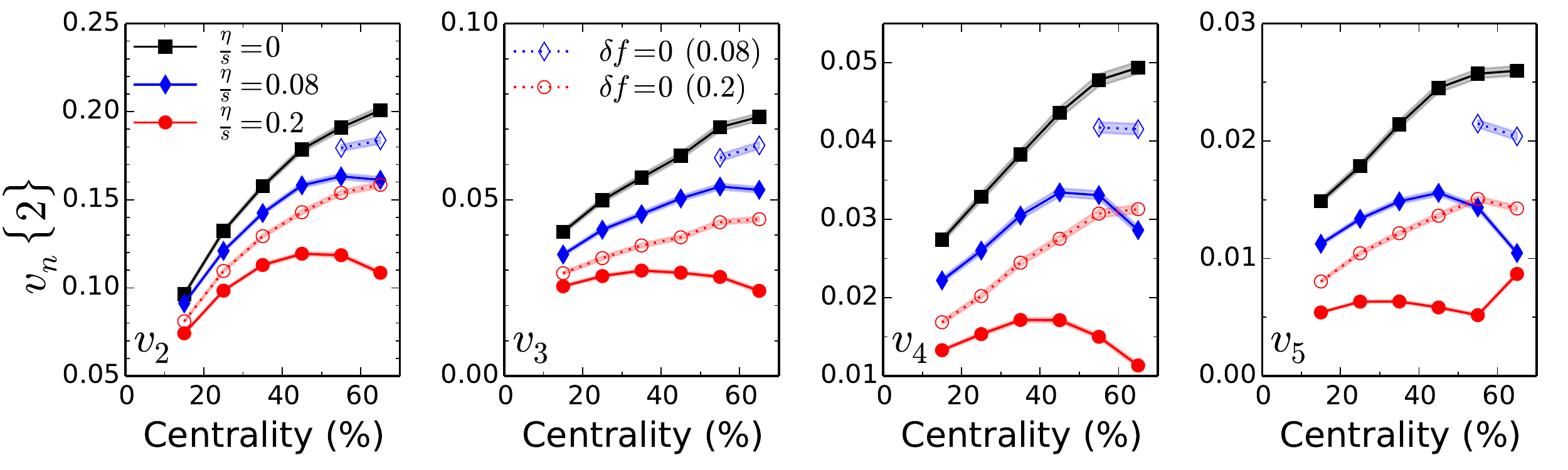}
    \caption{(Color online)
    \baselineskip=14pt
    $v_2$, $v_3$, $v_4$, and $v_5$ as function of collision centrality from ideal and viscous 
    hydrodynamics with KLN initial conditions. Each centrality bin contains 1000 events. Solid 
    black squares correspond to ideal hydrodynamics, while solid blue diamonds and solid red 
    circles correspond to viscous hydrodynamics with $\eta/s\eq0.08$ and 0.2, respectively.
    The open blue diamonds and red circles are also from viscous hydrodynamics with 
    $\eta/s\eq0.08$ and 0.2, respectively, but without including the viscous correction $\delta f$ at 
    freeze-out.
    \label{F5}
}
\end{figure*}
%
 
\begin{figure*}[!hbt]
    \includegraphics[width=\linewidth]{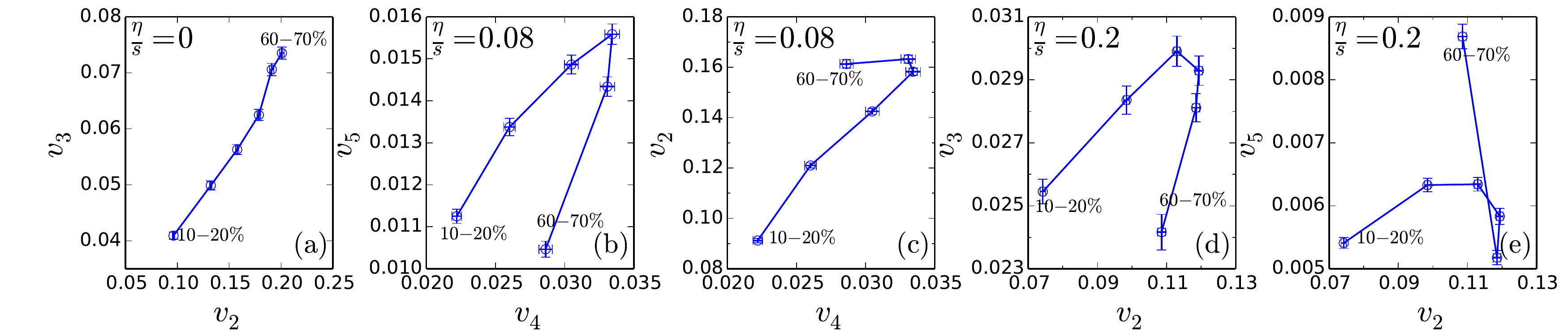}
    \caption{(Color online)
    \baselineskip=14pt
     Different types of $v_m-v_n$ correlation structures (``boomerang shapes'') from ideal and 
     viscous hydrodynamics with KLN initial conditions. Each centrality bin contains 1000 events. See
     text for explanation. 
     \label{F6}
}
\end{figure*}

As examples, we show in Fig.~\ref{F6} the centrality dependences of $v_n$, $n\eq2-5$, from ideal and viscous hydrodynamics, with and without including the viscous $\delta f$ corrections of the local distribution function at freeze-out. Most of these flow coefficients first increase and then decrease again as one moves from central to peripheral collisions, although a few simply increase monotonically over the range of centralities shown in the plot, and for the highest value of the shear viscosity studied here, $\eta/s\eq0.2$, the pentangular flow $v_5$ actually first rises, then drops, then rises again.\footnote{%
    \label{fn7}%
    \baselineskip 12 pt%
    Comparison of the open and filled circles in Fig.~\ref{F5} shows that this feature is actually caused by 
    a large viscous correction $\delta f$ at freeze-out. We found that the mechanism underlying the 
    appearance for $\eta/s\eq0.2$ of a dip of $v_5$ at $50\%-60\%$ centrality is that the inclusion of 
    $\delta f$ causes a jump by $\pi/5$ of the peak of the distribution of the relative angle between the 
    pentangular participant and flow planes, $\Psi_5{-}\Phi_5$. More generally, we observe in Fig.~\ref{F5}
    very large $\delta f$ corrections in peripheral collisions to all anisotropic flow coefficients. This raises
    doubts about the quantitative reliability of hydrodynamic simulations in the most peripheral bins and
    calls for an improved treatment in future work of the viscous corrections to the freeze-out distribution 
    function.
    } 
\newpage

The figure illustrates that several factors conspire to produce the observed centrality dependences of $v_n$: (1) the centrality dependences of the driving eccentricities $\epsilon_n$, (2) the centrality dependent duration of the hydrodynamic evolution which shortens the lifetime in peripheral collisions, cutting short the build-up of flow anisotropies, and which also controls the magnitude of the viscous corrections at freeze-out, and (3) the shear viscous damping of the anisotropic flow coefficients, especially for larger $n$. In peripheral collisions the viscous damping of $v_n$ is stronger than in central collisions \cite{Song:2007fn}, so as the shear viscosity increases the anisotropic flows peak at lower centrality, and this shift of the peak to lower centralities increases with harmonic order $n$.   

In Fig.~\ref{F6} we show that these trends can lead to qualitative changes in the shape of the correlation between $v_n$ and $v_m$ as function of centrality when we change the pair $(n,m)$ and/or the shear viscosity. Panel (a) shows a case where both $v_n$ and $v_m$ increase within a certain centrality range monotonically with centrality (here: $v_2$ and $v_3$ from ideal fluid dynamics). This type of correlation does not look at all like what is seen in the ATLAS data. A better representation of the experimental shape is obtained in panel (d) for $\eta/s\eq0.2$; however, the agreement is not perfect because in the hydrodynamic simulation $v_3$ peaks at smaller impact parameter than $v_2$ whereas the ``boomerang'' shape seen by ATLAS requires\break
\newpage\noindent 
that both $v_3$ and $v_2$ peak at the same centrality. We observe such a ``boomerang'' shape for the correlation between $v_5$ and $v_4$ for $\eta/s\eq0.08$, shown in panel (b) of Fig.~\ref{F6}. Panel (c) demonstrates that a ``boomerang'' with opposite orientation is obtained when both $v_m$ on the vertical axis and $v_n$ on the horizontal axis peak as a function of impact parameter, but $v_n$ then decreases faster than $v_m$ when the impact parameter is further increased. Such a situation is seen in the experimental data when ATLAS plots the correlation between $v_4$ on the vertical axis and $v_3$ on the horizontal axis (Fig.~12 in \cite{Aad:2015lwa}) but not reproduced by our hydrodynamic simulations (see Fig.~\ref{F5}b,c). Panel (e) of Fig.~\ref{F6} shows the very exotic form of the $v_m-v_n$ correlation that is possible when one of the two flow harmonics has a centrality dependence like the one seen for $v_5$ for $\eta/s\eq0.2$ in Fig.~\ref{F5} (right panel): in this case the ``boomerang'' can bend around to form a loop!

%
%

Let us summarize by concluding that the perfect ``boomerang''-shaped correlations between $v_2$, $v_3$ and $v_4$ observed by ATLAS in \cite{Aad:2015lwa} require that all three of these flow coefficients peak as functions of centrality at the same centrality. Figs.~\ref{F1} and \ref{F5} show that the purely hydrodynamic simulations with MC-Glauber and MC-KLN initial conditions and the choices of $\eta/s$ reported here are not able to reproduce this feature. This may be related to the large $\delta f$ corrections in peripheral collisions noted in footnote~\ref{fn7}, but a deeper study will be required to fully clarify this issue. 

\section{$q_n$ bins for event-shape selection}
\label{appa}

%
\begin{figure}[!hbt]
    \includegraphics[width=0.65\linewidth]{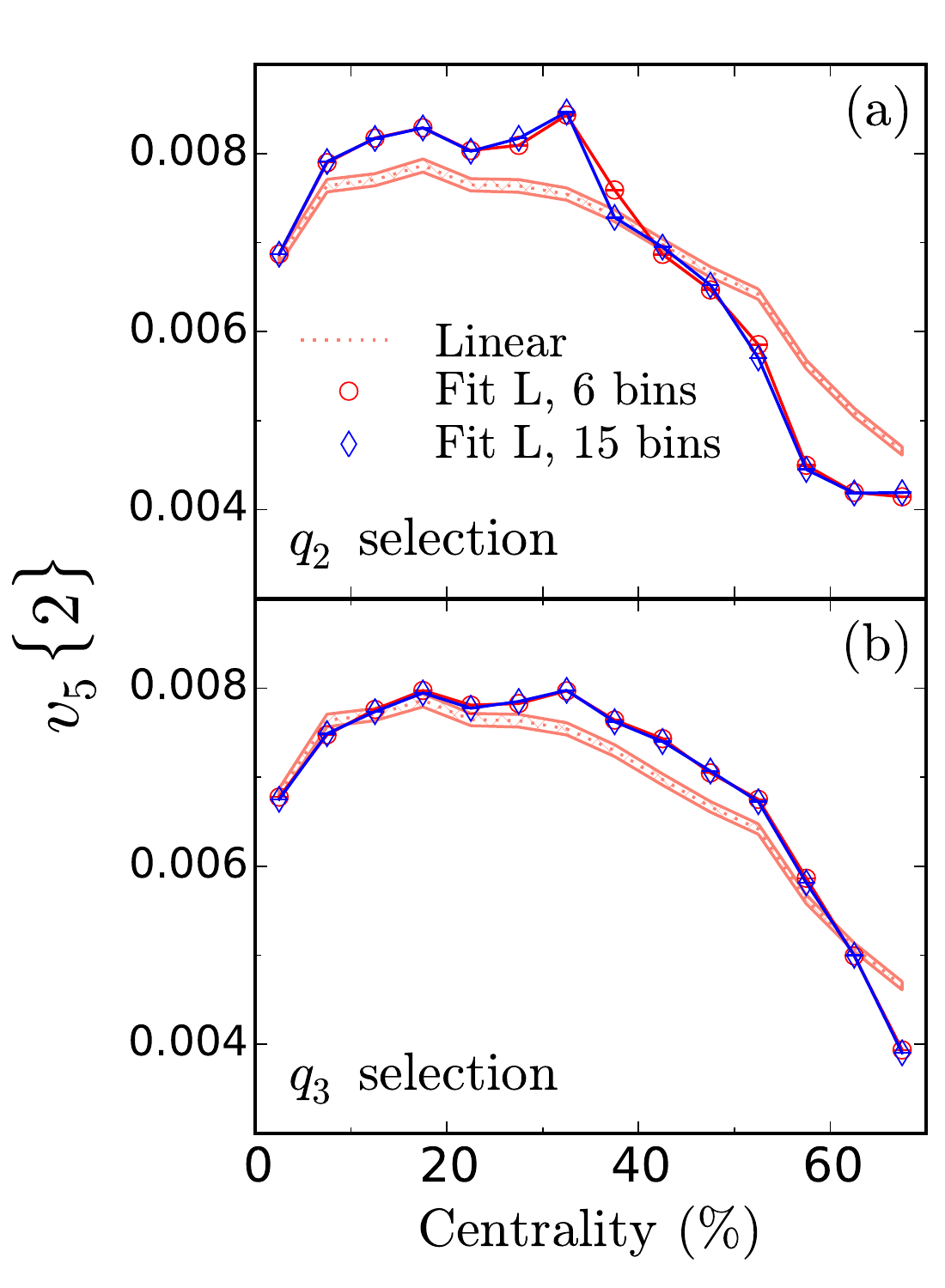}
    \caption{(Color online)
    \baselineskip=14pt
    The centrality dependence of the linear part of $v_5\{2\}$ obtained from the fit with Eqs.~(\ref{fit}),
    (\ref{v4v5_fit}), as shown in Fig.~\ref{F4}, but now compared for two different fits using 6 (red circles) 
    and 15 (blue diamonds) $q_2$ (a) or $q_3$ (b) bins when binning the 3000 events in each centrality 
    class by their event shapes. The red lines show the linear parts as defined in Eqs.~(\ref{v4v5}) for 
    comparison. Note the expanded vertical scale compared to Fig.~\ref{F4}.
     \label{F7}
}
\end{figure}
%
Due to event-by-event fluctuations, events with similar multiplicity may have quite different initial density distributions. The event-shape selection method was proposed to constrain initial event shapes by selecting certain events on the basis of their anisotropic flow coefficients \cite{Schukraft:2012ah, Petersen:2013vca, Huo:2013qma}. 

In ATLAS, events of a given collision centrality were subdivided into 15 $q_n$ bins (sometimes 14 $q_n$ bins when the two highest $q_n$ bins were combined) covering the following fractions of the $q_n$-ordered events \cite{Aad:2015lwa}: $0{-}0.01,\, 0.01{-}0.025,\, 0.025{-}0.05,\, 0.05{-}0.075,\, 0.075{-}0.1$, $0.1{-}0.2$, $0.2{-}0.3$, $0.3{-}0.4$, $0.4{-}0.5$, $0.5{-}0.6$, $0.6{-}0.7$, $0.7{-}0.8$, $0.8{-}0.9$, $0.9{-}0.95$, and $0.95{-}0.1$. In our case the number of events in each centrality class is much smaller than in the ATLAS experiment. Therefore, we use only 6 $q_n$ bins covering the fractions $0{-}0.1$, $0.1{-}0.2$, $0.2{-}0.5$, $0.5{-}0.8$, $0.8{-}0.9$, and $0.9{-}1$. In this Appendix we explore, as an example, the sensitivity of the decomposition of $v_5$ into its linear and nonlinear parts, as described in Sec.~\ref{sec3}, to the number of $q_2$ or $q_3$ bins used in the event-shape selection procedure.

In Fig.~\ref{F7} we show as circles and diamonds the linear parts of $v_5\{2\}$ as a function of collision centrality, computed by fitting in each centrality bin the dependence of $v_5\{2\}$ on $v_2\{2\}$ resp. $v_3\{2\}$ with the last equation in (\ref{fit}), using information from 6 or 15 $q_2$ resp. $q_3$ bins. One sees that in all except the 35\%-40\% centrality bin the fits converge to almost exactly the same $v_5^0$ values whether we use 6 or 15 $q$-bins. If we had used 15 $q$-bins in Fig.~\ref{F1}, the results would have looked rather noisy, due to statistical limitations. Fig.~\ref{F7} demonstrates that by reducing the number of $q$-bins to 6, in order to see the trends in Fig.~\ref{F1} more cleanly, we are not sacrificing accuracy when separating $v_n\{2\}$ into its linear and nonlinear contributions. 

With larger numbers of events than we have at our disposal, one can add additional $q_n$ bins near the two ends of the $q_n$ distribution, to better explore the full range of event shapes. Since the shape of the $q_n$ distribution depends on collision centrality, the optimal choice of the centers and widths of each $q_n$ bin may depend on centrality. With our limited statistics, we chose 6 bins whose positions and widths were distributed symmetrically around the median $q_n$. ATLAS \cite{Jia:2014jca} chose $q_n$ bins that were optimized to the Bessel-Gaussian distribution of $q_n$ in ultra-central collisions and whose positions and widths were asymmetric around the median $q_n$. They then used the same asymmetric binning scheme at all collision centralities even though the $q_2$ distribution first becomes more symmetric and then again asymmetric in the opposite direction as one proceeds from ultra-central to ultra-peripheral collisions. Fig.~\ref{F7} shows that, for the purpose of separating linear and nonlinear contributions to the higher-order flow harmonics the precise placement of the $q_n$ bins does not matter. 


\end{document}